# Electronic and Optically controlled Bi-functional Transistor based on Bio-Nano Hybrid Complex

Vikram Bakaraju[1], E. Senthil Prasad[2], Brijesh Meena[3,4], and Harsh Chaturvedi[3, *]

[1]*Department of Physics, University of Antwep, Antwep, Belgium*

[2]*Institute of MicrobioTechnology (IMTECH), Chandigarh, India*

[3]*Center for Energy, Indian Institute of Technology (IIT), Guwahati, Assam, India*

[4]*G Lab Innovations Pvt. Ltd., Kolkata, West Bengal, India*

*harshc@iitg.ac.in

**Abstract**:

We report optically as well as electronically controlled, bio-electronic field effect transistor (FET) based on the hybrid film of a photo-active purple membrane and electronically conducting single-walled carbon nanotubes (SWNTs). Two dimensional (2D) crystal of bacteriorhodopsin forms the photoactive center of the bio-nano complex, whereas one dimensional (1-D) pure SWNTs provides the required electronic support. Hybrid structure shows conductivity of 19 μS/m and semiconducting characteristics due to preferential binding with selective diameters of semiconducting SWNTs only. The bio-electronic transistor fabricated using direct laser lithography shows photoconductivity of 13.15μS/m and well controlled optical and electronic gating with significant On/Off switch ratio of 8.5. The device demonstrates well-synchronized electronic and optical gating wherein; n-type FET shows complimentary p-type characteristics due to optically controlled 'proton-pumping' by bacteriorhodopsin. Fabricated bio-electronic transistor displays both electronically and optically well controlled, bi-functionality.

**Introduction**:

Bioelectronics aim to use electronic/optically active, functional biological molecules (chromophores, proteins, etc.) as an active material for electronic or photonic devices. Research on bio-nano hybrid materials is being actively pursued for potential application in developing functional, electro-optical devices and biological sensors.(Bräuchle, Hampp et al. 1991, Hampp 2000, Huang, Wu et al. 2004) Donor acceptor system, based on SWNTs functionalized with inorganic, organic polymers,(Borghetti, Derycke et al. 2006) biological DNA,(Shim, Shi Kam et al. 2002, Keren, Berman et al. 2003, Staii, Johnson et al. 2005) protein etc. have been widely reported in pursuit of fabricating such hybrid devices (Singh, Pantarotto et al. 2005). Devices such as field effect transistors, sensors, rectifiers based on SWNTs functionalized with optically active materials like quantum dots, inorganic ruthenium dyes or optically active molecules has been fabricated.(Katz and Willner 2004) Photoactive proteins/molecules can bind either covalently or non-covalently with SWNTs, to form stable donor-acceptor system.(Guldi, Rahman et al. 2005, Sharma, Prasad et al. 2015) However, non-covalent functionalization of SWNTs is generally preferred for electro-optical devices, as it preserves the electronic nature of SWNTs in the functionalized hybrid complex.

Diverse devices are being proposed, based on SWNTs functionalized with optically active, biological molecules.(Barone, Baik et al. 2005, Wang 2005) Bacteriorhodopsin (bR) is a stable optically active protein, widely proposed for its technological application in developing various electronic and photonic devices such as optical data storage, electro--optic memory, logic gates, photo-chromatic and holographic systems. Recent reports of functionalization of Bacteriorhodopsin with nanoparticles (quantum dots, nanotubes) shows active interest in using functionalized bio-nano hybrid complex for developing novel photovoltaic, electro-optic devices.(Jin, Honig et al. 2008, Lu, Wang et al. 2015) Here in, we report fabrication of electronic and optically bifunctional, field effect transistor (FET) based on SWNTs functionalized with bacteriorhodopsin. Fabricated bio-electronic transistor shows intrinsically semiconducting electro-optical properties with well controlled electronical and optical gating, optical doping and photoconductivity.

Bacteriorhodopsin forms the optical centre of the Purple membrane (PM). Purple membrane is made up of 75% protein and 25% lipid.(Lozier, Bogomolni et al. 1975) Only protein present in the purple membrane is the bacteriorhodopsin, which acts as a light driven proton pump.(Lozier, Bogomolni et al. 1975) Photo-cycle of bacteriorhodopsin in the purple membrane has been well characterized and duly reported. PM remains stable up to 80 C, has buoyant density of 1.18 g/cm$^3$, Refractive index of 1.45 – 1.55 and its natural stable crystalline structure makes purple membrane excellent two dimensional (2D) optical material for the development of bio-electronics.(Xu, Bhattacharya et al. 2004, Wang, Knopf et al. 2006) Optoelectronic and photonic applications demand high quantity (several milligrams) thin films of purple membrane. For large-scale incubation of bacteriorhodopsin, Halobacterium cultivation was performed in a 7 litre photo-bioreactor. Purified purple membrane were characterized using UV-Vis spectrophotometer by taking ratio of absorptions at 280 nm and 560 nm. Along with high yield of 14.4 mg/l, extracted PM shows high quality estimated using absorption peaks, to be around ~ 2.0 to 2.1. (**Figure 1**) 2 mg/ml of this extracted, protein aliquot was used for further functionalization with SWNTs.

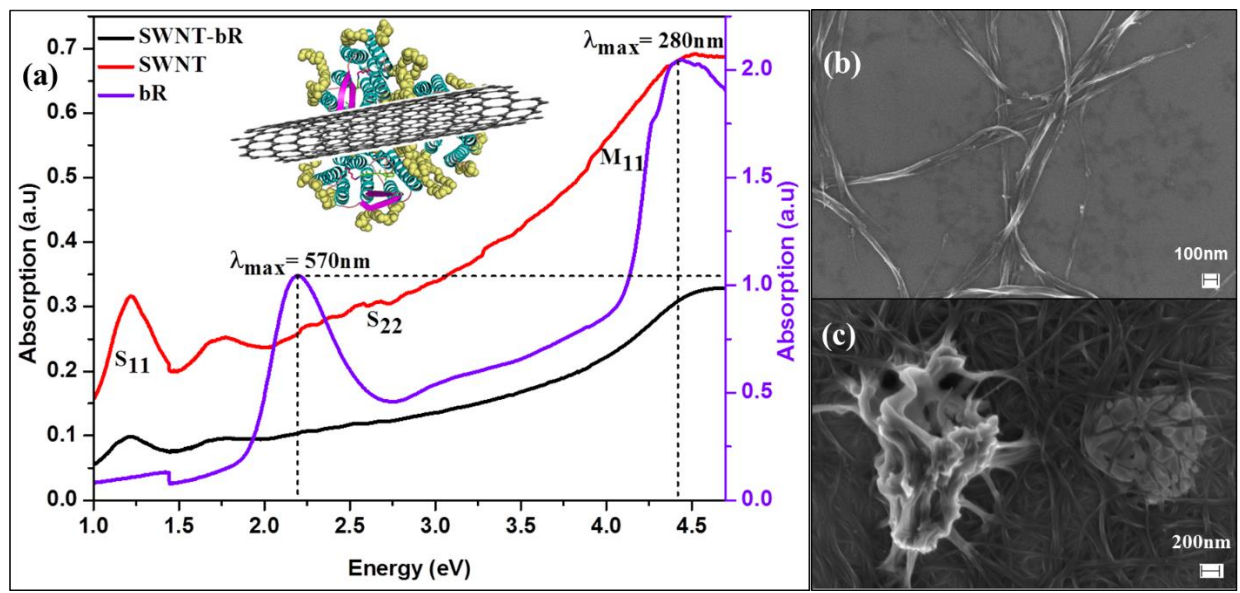

Figure 1 (a) Absorption spectra of pure SWNT (red) and bacteriorhodopsin (bR) (blue) control solutions and that of functionalized SWNT –bR complex. (b) SEM image of pure SWNT (c) SWNT functionalized with bR complex.

Proposed Bio-nano devices based on SWNTs are limited due to inherent hydrophobic nature of pristine SWNTs. However, aqueous solutions of SWNTs have been reported using surfactants.(Vaisman, Wagner et al. 2006) Here in, biocompatible phospholipid polyethylene glycol amine was used as the surfactant to prepare well dispersed aqueous solution of pure SWNTs. (Liu, Tabakman et al. 2009)

**Materials and Methods**:

Ultrapure (99.5%) SWCNTs, purchased from NanoIntegris in the form of sheets (batch number: P10-126) and were dispersed in water using 1,2-Distearoyl-*sn*-Glycero-3-Phosphoethanolamine-*N*-Amino [(polyethylene Glycol) 2000] (Ammonium Salt) (Product code: 880128P), bought from Sigma Aldrich. Well dispersed stable aqueous solution of SWNTs was obtained for the ratio of 1:7.5 (1 part of SWCNT with 7.5 parts of PlPEG amine). Measured absorption spectra (UV-vis NIR) of the prepared aqueous solutions, do not show any discernible changes in the concentration of SWNTs, over weeks. Aqueous solution of $10^{-2}$M concentration of synthesized Purple membrane (PM) and dispersed SWNT solution was prepared. Aggregated SWNT functionalized with PM were optically aggregated and then separated by the established protocol as reported.(Sharma, Prasad et al. 2015) Aqueous solution of SWNTs stirred with purple membrane were illuminated by broadband mercury lamp for three hours (3 hrs) and aggregated flocs of functionalized SWNTs were separated from the dispersed supernatant using centrifugation. Separated aggregates of functionalized SWNTs were then drop casted onto the fabricated device and further characterized using Raman spectroscopy. Figure 1b shows SEM images of pristine (top) as well as that of separated aggregates of SWNT functionalized with bacteriorhodopsin. Bacteriorhodopsin crystal in purple membrane is expected to be ~ 500 nm; whereas pristine SWNTs are ~ 1-2 micron in length with diameters varying from 0.7 nm up to 1.8 nm. HR-SEM images have been taken using ZIESS system with back scattering configuration. Images with SWNTs extended out from the matrices of two dimensional bacteriorhodopsin-protein complex, particularly displays surface functionalization and stable hybrid complex required for further fabrication of functional devices.

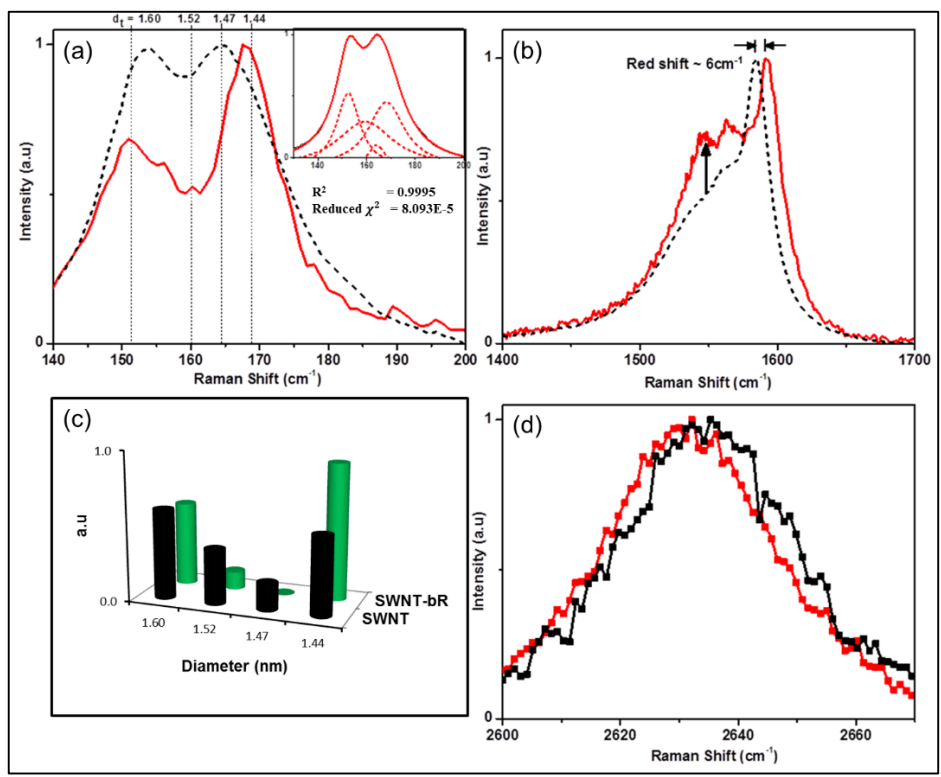

Figure 2: Raman spectra showing (a) RBM of SWNT (black) and functionalized complex of SWNT-bR(red) showing stable preferential binding with specific diameters. (b) G band showing changes in $G^+$ (1591 cm$^{-1}$) and $G^-$ (1540cm$^{-1}$) peak of SWNT-bR Complex as compared to pure SWNTs (c) Histogram shows preferential binding, enrichment SWNTs of specific diameters (d) G' band shows slight red shift (3cm$^{-1}$).

Hybrid complex of SWNT functionalized with bacteriorhodopsin was further characterized using Raman spectrometer with 632 Red laser line. Raman spectra (**Figure 2**) of the functionalized samples show stable and preferential binding of bR with SWNTs of specific diameters. RBM of Raman spectra essentially depends on the diameter of the SWNTs which may be calculated using the relation $\omega_{RBM} = (\alpha_{RBM}/d) + \alpha_{bundle}$. Where, $\alpha_{RBM}, \alpha_{bundle}$ are constants and $d$ is the diameter of the SWNT corresponding to the RBM peak frequency $(\omega_{RBM})$.(Bachilo, Strano et al. 2002)

Histogram (Figure 2 c) as calculated from RBM of Raman spectra (Figure 2 a) shows enrichment of specific diameters of SWNTs in the hybrid complex. Discernible changes in G band of functionalized SWNT is observed as compared to pure SWNTs. Changes in G$^-$ band are related to metallicity of the SWNTs and have been shown to vary with the changes in the electronic properties of the nanotubes. Red shift of ~ 6 cm$^{-1}$ is observed in SWNTs functionalized with the BR. This red shift in the functional SWNTS can be attributed to strong interaction and electronic doping by the bacteriorhodopsin.(Rao and Voggu 2010) Using similar approach as reported by Rao et al to estimate electronic doping by corresponding shifts in the Raman peaks, red shift in our devices indicates doping by estimated charge density of about n ≈ $3*10^6$ cm$^{-2}$.

Devices were fabricated using standard lift off technology and direct laser lithography to pattern metallic pads on p-type doped silicon wafer (purchased from Semiconductor Wafer. Inc.). Metallic patterns act as source and drain whereas sample holder (chuck) acts as back gate of the field effect transistor. Hybrid film of the purple membrane and SWNTs acts as the functional semiconducting channel of the fabricated FET.

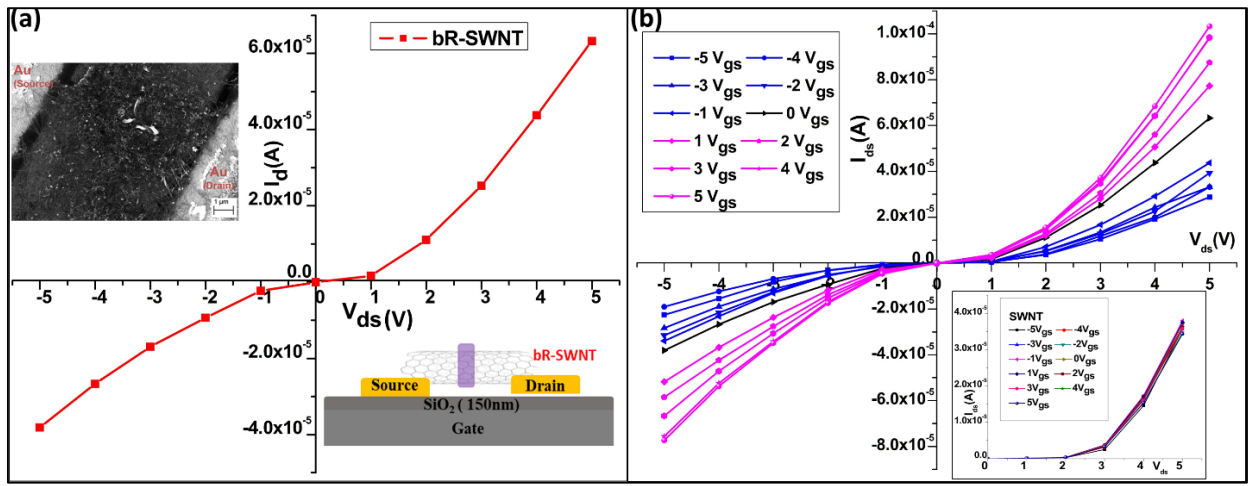

Figure 3: (a) Comparative I-V characteristics of FET based on SWNTs functionalized with bacteriorhodopsin (bR). SEM image of device fabricated is shown as an inset along with the schematic of the FET. (b) I-V characteristics of the fabricated FET based on functionalized SWNT-bR complex for a range of gate bias applied [-5$V_{gs}$, +5$V_{gs}$].

Figure 3 (a) (inset) shows the HR SEM image of the fabricated FET. Devices were electro-optically characterized using integrated probe station having micromanipulators attached with a dedicated Keithley 4200 SCS Semiconductor characterization system. Control I-V measurement of the devices which were fabricated using pristine SWNTs do not exhibit any significant gate control (insert, Figure 3 (b)). However, devices functionalized with bR show characteristic n-type semiconducting behaviour and significantly enhanced control on current using gate voltages. FET shows higher (maximum observed ON/OFF ratio: 8.5) current for positive gate voltages (0 to 5 V) as compared to applied negative voltages (0 to -5 V). The decrease in current is observed with corresponding increase in applied negative gate voltages (33 µS/m for $V_g$= 5V, 9.6 µS/m for $V_g$= -1V and 0.14 µS/m for $V_g$= -5V) demonstrating well controlled 'Electronic Gating' of the device.

Moreover, under light, "optical doping" is observed. Figure 4 compares the electronic characteristics of the device under dark and broadband illumination by mercury lamp. The FET devices based on SWNTs functionalized with bR shows n-type characteristics under dark condition but under broadband illumination from mercury lamp, FET devices switches into p-type characteristics. Optical doping is observed in the hybrid device due to proton charge transfer from photoactive 'proton pump' bR to SWNT (Keiichi Inoue, 2016). Figure 4 (a) demonstrates I-V characteristics of the Functionalised FET under light and dark conditions. Device is 'On' for positive gate voltages (0 to 5 $V_{gs}$) in dark condition and is 'On' for negative gate voltages (-5 $V_{gs}$ to 0) in light condition (figure 4). On/Off ratio of 8.5 in dark condition and 4.9 in light condition is observed. The device shows maximum Photocurrent (ILight – IDark) of 40 μA at 5 Vds and 0 Vgs with the conductivity of 13.15μS/m (insert Figure 4 (b)). Comparing the current between both light and dark conditions, it is seen that current decreases by 60% (5 $V_{ds}$ & 5 $V_{gs}$) for positive gates and increases by 300% (5 $V_{ds}$ & -5 $V_{gs}$) for negative gates under light.

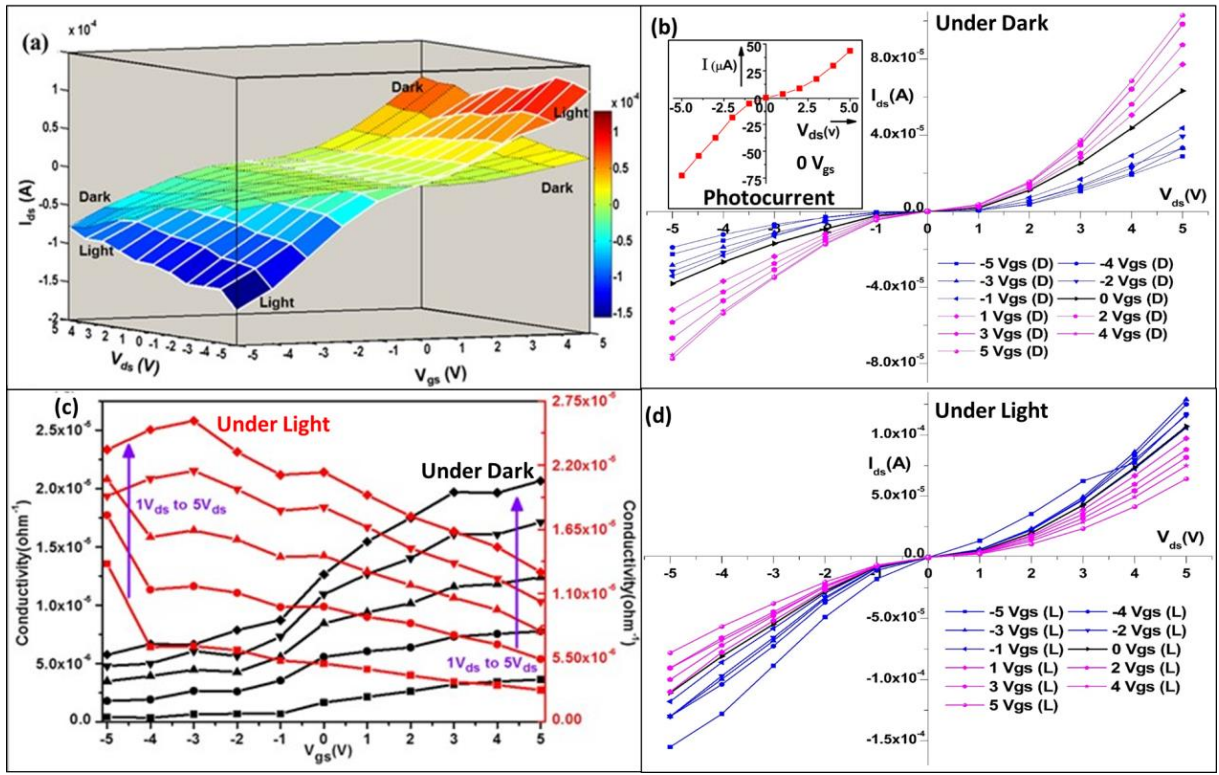

Figure 4: (a) 3-Dimension plot of current-voltage (I-V) response of the device with varying drain –source and gate voltages applied under conditions of dark and light. (b) I-V plot of the bR-SWNT FET at various gate volatges under dark condition (pink for positive gate volategs and blue for negative gate voltages). Insert shows the photocurrent at 0 gate voltages. (c) Plot showing "Optical

Doping" with device switching from n-type (black) under dark to p-type under light (red). (d) I-V plot of the bR-SWNT FET at various gate voltages under Light condition.

Thus, we observe singinificant "Electronic Gating" of the functionalized FET, and excelent "Optical Doping" under light and dark conditions. This can be attribute to well-established photocycle and proton transfer properties of bR, there by affecting electronic conduction through the functionalized SWNT. SWNT FET devices works on the principle of schottky barriers at the metal semiconductor contact (Campbell, 2011). SWNTs are generally considered to be ambiopolar, with symmetrical changes in current for both positive and negative gate voltages. However, since electronic transport through the device significantly depends on the Schottky barrier; any modulation in electronic concentration is expected to cause significant changes in the electronic property of the device. Although, further research is in progress to understand the charge transfer and transport mechansim in this hybrid donoer-acceptor system, the results confirm optical activation of the 'proton – pump' purple memberane with significant charge transfer onto the SWNTs, there by converting n-type functionalized SWNT FET into p-type devices.

In conclusion, we have shown stable, optically functional bioelectronic device based on purple memberane and SWNT complex. Both "optical doping" and significant "optical switching" is observed in fabricated FETs. Devices also show considerable photocurrent and electronic gating. Well-controlled electro-optical functionality of the device is due to strong interaction and charge transfer between the 2-dimensional optically active bacterirhodopsin and electronic single walled carbon nanotubes, as also affirmed by the Raman spectral analysis. We believe the results discussed here will be important for diverse biophotonic, bioelectronic, biosensing and photo-voltaic applications. Results shown here may also contribute in realizing functional reconfigureable bio-photonic, electro-optical devices with devices showing reversible and controlled change in the type of majority carriers, depending on the condition wether the device is under illumination or is in dark. "Optical doping", "Electro-Optical switching" and "Electronic gating" – the fact that each of these phenomenon reported above are using bio-nano hybrid complex and realized in the same device, promises potential for widespread application in diverse areas from biological sciences to electronic photonic hybrid sensors and devices.

**Acknowledgements:**

Authors are deeply indebted to the funding agencies DST (DST/TSG/PT/2012/66), Nanomission (SR/NM/NS-15/2012) for generous grants and support. Part of the work was done at IISER Pune.